# Anticipated impacts of Brexit scenarios on UK food prices and implications for policies on poverty and health: a structured expert judgement update

Martine J Barons[1] [0000-0003-1483-2943]• Willy Aspinall[2] [Orcid 0000-0001-6014-6042]

[1]Director of the Applied Statistics & Risk Unit Applied Statistics & Risk Unit, Department of Statistics, University of Warwick, Coventry, UK. Martine.Barons@warwick.ac.uk Corresponding author.

[2] Principal Consultant, Aspinall and Associates, Tisbury, and University of Bristol, Bristol, UK
Willy.Aspinall@bristol.ac.uk

## Abstract

**Introduction** Food insecurity is associated with increased risk for several health conditions and with increased national burden of chronic disease. Key determinants for household food insecurity are income and food costs. Forecasts show household disposable income for 2020 expected to fall and for 2021 to rise only slightly. Prices in the same periods are forecast to rise. Thus, future increased food prices would be a significant driver of greater food insecurity.

**Objectives** To investigate food price drivers for household food security and its health consequences in the United Kingdom under three different agreement scenarios for the end of the transition period for Britain's exit from the European Union, in December 2020. To estimate the means, medians and $5^{th}$ percentile and $95^{th}$ percentile quantiles of projected food basket price distributions by April 2022, which would result from each agreement scenario.

**Design** Structured expert judgement elicitation, a well-established method for quantifying uncertainty, using experts. In July 2020, each expert estimated the median, $5^{th}$ percentile and $95^{th}$ percentile quantiles of changes in price to April 2022 for ten food categories under three end-2020 settlement Brexit scenarios: A: full WTO terms; B: a moderately disruptive trade agreement (better than WTO); C: a minimally disruptive trade agreement. These scenario price quantiles are enumerated from the judgements of a group of experts, aggregated with performance scores based on the statistical accuracy and informativeness of the experts, as determined with a set of calibration items.

**Participants** Eight specialists with expertise in food procurement, retail, agriculture, economics, statistics and household food security.

**Results** When combined in proportions used to calculate Consumer Prices Index food basket costs, the median food price change under full WTO terms is expected to be +17.9% [90% credible interval:+5.2%, +35.1%]; with moderately disruptive trade agreement: +13.2% [+2.6%, +26.4%] and with a minimally disruptive trade agreement +9.3% [+0.8%, +21.9%].

**Conclusions** The number of households experiencing food insecurity and its severity are likely to increase because of expected sizeable increases in median food prices in the months after Brexit, whereas low income group spending on food is unlikely to increase, and may be further eroded by other factors not considered here (e.g. COVID-19). Higher increases are more likely than lower rises and towards the upper limits, these would entail severe impacts. Research showing a low food budget leads to increasingly poor diet suggests that demand for health services in both the short and longer term is likely to increase due to the effects of food insecurity on the incidence and management of diet-sensitive conditions.

# Introduction

Food insecurity, the lack of access to sufficient nutritious food, is associated with multiple negative outcomes including diet-sensitive chronic diseases. An important driver of household food insecurity is the costs of food and other essentials relative to incomes. In 2016, the United Kingdom (UK) voted to relinquish its membership of the European Union (EU), known colloquially as 'Brexit', and the transition period is due to end on 31st December 2020. UK reliance on food imports, including from EU, is significant and food price rises have been widely forecast [1].

In March 2020, we published the results of an expert elicitation undertaken in July 2018 for food prices to June 2020, under the key assumption that Brexit would take place on 29th March 2019, as originally intended [2]. Subsequently, eight of the same experts provided updated estimates, between March and August 2020, taking into account the changed Brexit delivery timescale but excluding the confounding impact of COVID-19 on food supply. This is reasonable since the food supply shortages seen at the beginning of the UK lockdown resulted from the inability of the UK just-in-time food supply to cope with the uptick in demand driven by the lockdown, when households stocked up for the possibility of self-isolation due to COVID19 [3] and spent more on eating at home, instead of using restaurants and other outside food providers.

# Methods

In the updated analysis, we chose three possible Brexit scenarios instead of the two ("deal" or "no-deal") in the original investigation. This is to reflect the change of political landscape, including the fact that the UK has now formally left the EU and is in a transition period, negotiating alternative trade arrangements. The three Scenarios are: A: full WTO terms, (broadly similar to no-deal); B: a non-specific but moderately disruptive trade agreement (better than WTO); C: a minimally disruptive trade agreement (broadly similar to the arrangements under EU membership). We also asked our panel of experts for their views on the relative likelihood of each of these three scenarios being realised.

The expert elicitation was carried out using the same method as before [2]. The experts' estimates were combined under the Cooke protocol and the combined estimates of food price changes in the 10 categories in the CPI calculation, under each of the three scenarios, are shown in Table 1. We estimated the cost changes for the CPI basket and the 'healthy baskets' from MacMahon & Weld [4] for a family of four and for a single pensioner, containing slightly differing proportions of food category items in each. We updated the baseline basket costs to July 2020, using ONS historic inflation data [5]. These July 2020 costs were then projected to end December 2020 using a Gaussian distribution derived from the last five years' July-to-December overall food price change data. Subsequent food price changes, from December 2020 to April 2022, were calculated from the elicitation findings via the Bayesian Belief Network (BBN) model we used previously [2].

# Results

The different food category prices show different patterns of future change under the three Brexit scenarios which our experts considered.

Three food categories (Coffee, Tea & cocoa; Soft drinks etc.; Sugar, jam, etc.) are expected to have similar price rises under any of the scenarios.

Three food categories (Vegetables; Fruit; Milk, cheese & eggs) show substantial differences across the scenarios, with the minimally disruptive scenario producing smaller median price rises of around +5%. Given these are important food groups for health and that even a +5% price rise can be significant for those on low incomes, and for a large number of households, this may produce adverse long and short term national health outcomes, as described in [2]. The moderately disruptive Scenario B shows median price rises around +10% to +15% for these categories and the full WTO terms Scenario A shows median price rises around +16% to +24%. The higher the price rises on essential, staple foods, the greater the number of households that will be affected.

Three food categories (Oils & fats; Fish; Bread & cereals) show a pattern where the difference between median price increases under Scenario C versus Scenario B are more substantial than between Scenario B and Scenario A. That said, in each case, the 90% credible interval under Scenario C contains the medians of Scenarios B and A. Fish and Bread & cereals are also very important food groups for health, and the same comment applies as for Vegetable, Fruit and dairy, such that even the smallest median increases could have substantial impacts of the lowest income households.

Meat is the category showing the highest median price increase under Scenario C and this category cost also rises slightly under the other two scenarios. In many diets, meat is an important source of protein, and a price increase on the scale of Scenario C is likely to lead to a move towards lower-quality meat for lower-income households. It is possible that it will also accelerate moves towards vegetarian diets, although the taste of meat and animal products is enough of a barrier to prevent dietary change for most people; the principal motivation given by those who do change dietary regime is not cost but for personal reasons, such as health [6].

Note that the elicitation protocol, at individual expert level, does not entail any consistency checks for the scenario price changes per food category. There will, inevitably, be some jitter in joint price change estimates when individual judgements are combined in the weighted group solutions. Thus, differences of order +/- 1% in the findings are not meaningful. Also, the distribution support for each food type, provided by the experts, is predominantly for some measure of price increase and, in all cases, decreases are relatively low probability events; it should not be surprising that, for many foods, the combined distribution is heavily positive. Thus, while individual food prices may not change dramatically, the compound effect in a basket of foods can be substantial.

The weighted combination of the three alternative exit agreement scenarios shows projected mean and median percentage CPI changes (+13.6% and +13.3% respectively, Table 1), very similar to those for Scenario B, confirming the latter represents a middle-of-the-road Brexit scenario, as far as anticipated subsequent changes in food prices are concerned. The weighted combination 90% credible interval percentage CPI basket change spread [+6.1%, +22.4%] is narrower than individual scenario spreads. Counterintuitively, but as proven with judgements by individual experts, combining alternative scenario distributions jointly in this way can produce price change forecasts whose quantiles are more informative than those of the separate, contributing individual expert distributions. The costs in £ of the CPI basket, healthy family of four and healthy pensioner baskets are given in Table 1.

The CPI basket reflects typical spending across all household types and income bands. Under Scenario C, this is expected to rise by a mean of +10.0%, leading to a median weekly price increase of +£8.44, 90% credible interval [+£1.14, +£18.54]. Under Scenario B, this rises to mean increase of +13.6%, median +£12.58 [+£2.59, +£23.05] and, under Scenario A, to a mean increase of +18.7%, median +£16.29[+£5.41, +£29.36].

| | Food category percentage price changes by April 2022 median (5$^{th}$, 95$^{th}$ percentile) |
|---|---|

| CPI category | A: full WTO terms | B: moderately disruptive trade agreement | C: a minimally disruptive trade agreement. |
|---|---|---|---|
| Soft drinks etc. | 9 (-1, 37) | 9 (-2, 30) | 9 (-1, 30) |
| Coffee, tea & cocoa | 6 (-1, 34) | 5 (-2, 30) | 6 (0, 31) |
| Sugar, jam, etc. | 6 (-7, 25) | 5 (-8, 24) | 5 (-5, 20) |
| Vegetables | 16 (-8, 51) | 10 (-4, 30) | 5 (-5, 16) |
| Fruit | 24 (-10, 61) | 14 (-7, 40) | 5 (-8, 24) |
| Oil & fats | 20 (-8, 47) | 20 (-10, 38) | 5 (-11, 27) |
| Milk, cheese & eggs | 17 (-6, 50) | 10 (-8, 21) | 5 (-5, 19) |
| Fish | 19 (-5, 44) | 18 (-4, 31) | 10 (-4, 29) |
| Meat | 20 (0, 57) | 18 (-1, 30) | 17 (-1, 40) |
| Bread & Cereals | 19 (0, 40) | 16 (-4, 34) | 5 (-5, 19) |
| | | | |
| **Overall projected % change in ONS CPI sub-Foods, with category weights** | **Mean +18.7% ±9.2 Median +17.9% [+5.2, +35.1]** | **Mean +13.6% ± 7.3 Median +13.2% [+2.6, +26.4]** | **Mean +10.0% ± 6.5 Median +9.3% [+0.8, +21.9]** |
| | Food basket cost changes by April 2022 (in £'s) | | |
| **Change in CPI weekly cost relative to 2020 year end Basket total £59.35*** | **Mean +£16.70 ± £7.33 Median +£16.29 [+£5.41, +£29.36]** | **Mean +£12.64 ± £6.19 Median +£12.58 [+£2.59, +£23.05]** | **Mean +£8.90 ± £5.31 Median +£8.44 [+£1.14, +£18.54]** |
| **Change in family of 4 Healthy Food Basket basis weekly cost £95.41**** | **Mean +£20.00 ± £10.04 Median +£19.04 [+£5.30, +£38.16]** | **Mean +£14.35 ± £7.76 Median +£13.74 [+£2.91, +£28.34]** | **Mean +£10.81 ± £7.22 Median +£10.05 [+£0.58, +£23.93]** |
| **Change in single pensioner Healthy Food Basket basis weekly cost £35.92**** | **Mean +£7.10 ± £3.71 Median +£6.73 [+£1.66, +£13.84]** | **Mean +£5.14 ± £2.88 Median +£4.89 [+£0.93, +£10.42]** | **Mean +£4.04± £2.63 Median +£3.78 [+£0.30, +£8.82]** |
| Agreement scenario -- weighted alternatives | | | |
| **Overall % change ONS CPI sub-Foods, with category weights** | **Mean +13.6% ±5.0 Median +13.3% [+6.1, +22.4]** | | |
| Food basket total costs by April 2022 (in £'s) | | | |
| **Projected CPI weekly Basket cost (inflation adjusted 2020 year end cost £59.35*)** | **Mean +£71.70 ± £4.14 Median +£71.53 [+£65.28, +£78.82]** | | |
| **Projected cost family of 4 Healthy Food Basket (inflation adjusted 2020 year end cost £95.41**)** | **Mean +£109.93 ± £5.42 Median +£109.52 [+£101.84, +£119.48]** | | |
| **Projected cost single pensioner Healthy Food Basket (inflation adjusted** | **Mean +£41.16 ± £1.99 Median +£41.01 [+£38.18, +£44.67]** | | |

| **2020 year end cost £35.92\*\***) | |

For the healthy basket for a family of four, the median increase under Scenario C is £10.05 per week, 90% credible interval [£0.58, £23.93]. Under Scenario B the median increase is £13.74 [£2.91, £28.34] and under Scenario A, £19.04 [£5.30, £38.16].

For the healthy basket for a single pensioner, the median increase under Scenario C is +£3.78 per week, 90% credible interval [+£0.30, +£8.82]. Under Scenario B, the median increase is +£4.89 [+£0.93, +£10.42] and under Scenario A, £6.73 [£1.66, £13.84].

*Table 1 \*Based on Office for National Statistics Table A2 2018year-end data (March 2018): selected basket subfood category weekly costs; total for the 10 items, updated using ONS data to July 2020 and projected to December 2020. \*\*Based on MacMahon and Weld Northern Ireland minimum essential healthy basket subfood category weekly costs at November 2014 Tesco prices, updated using ONS data to July 2020 and projected to December 2020. For two adults and two children, one in preschool (aged 2–4) and one in primary school (aged 6–11), total cost for the 10 items=£95.41; for a single pensioner, the corresponding selected items cost=£35.92.*

To illustrate some plausible futures beyond the median outcome, suppose that under Scenario B (as middle-of-road) the costs of all foods in that basket were to rise each to lie in the range above its own 84th percentile percentage price change value, the weekly CPI basket cost increases by +£31.57, i.e. the weekly CPI basket total cost would go to £90.92. This contrasts with the expected weekly CPI basket cost (£59.35) increase of +£12.64 (Table 1).

Under Scenario A, median CPI basket price increase is +£16.29 (Table 1). If Fruit and Vegetables only rise to their 95th percentile values, and all other prices remain at the median, the CPI basket price increase is Mean +£19.89 ± £6.84 Median +£19.50 90% CI [+£9.31, +£31.70]. (Note the new 90% credible interval is marginally smaller as a result of 'better' uncertainty information on Fruit and Vegetables).

Compared to the earlier estimates of food costs under the previous Brexit 'deal' and 'no-deal' scenarios [2], the updated elicitation shows that Scenario C is likely to yield higher food price rises than under the Brexit 'deal' scenario, had the UK had left the EU on 29th March 2019. In contrast, the price increase estimates for the current Scenario A are slightly lower than the previous estimates for a Brexit 'no-deal', and the credible interval is narrower. This said, it is notable that in this updating elicitation, overall food prices are not expected to fall under any scenario.

## Discussion

Food is a substantial portion of household expenditure, especially in low-income households. Falling incomes and rising prices in general, including food prices, will lead to a need for many to reduce food purchasing in order to meet other essential living costs. The corollary is that food purchasing would be further pushed toward lower cost, nutrient poor items, driving more households into food insecurity.

From this updating elicitation, and on a timescale of fifteen months after the end of the Brexit transition period, food price rises are likely to be significant and, with time, may become substantial; moreover, such increases could start to be felt by the whole population very quickly [1]. To compound matters, incomes are expected to stagnate or fall over the same period. A comparison of independent forecasts for the UK economy [7] shows a net fall in median household real disposable income of about 2.8% for the six months of 2020 that remained after the elicitation was completed. Such estimates range from falls of 1.1% to 5.8%.  Forecasts for 2021 range from a rise in disposable income of +4.7% to a fall of -2.0%, with a median +1.1% rise.  These same forecasts project rises on consumer prices index from -0.01% to +1.8% (median +0.4%) for 2020, and from +0.5% to +3.2% (median +2.0%) for 2021. In other words, any economic recovery is likely to be out-paced by food price ranges over those timeframes.

After housing and childcare costs, the highest category of household expenditure in Northern Ireland is on a minimum essential 'healthy' food basket [4]. Recent research estimates that the UK Global Tariff alone, which will apply wherever a trade deal has not been agreed, will add £246.38 per year to the food bill for a family of four. Cost increases due to exchange rate changes and 'on costs' due to transport delays, meeting border infrastructure costs and import processing bureaucracy are additional [8].

The North East, West Midlands and East of England have been assessed as UK regions most likely to suffer negative economic impacts because of Brexit, as they are disproportionately reliant on the most exposed sectors for output and employment [9].

On the basis of our updating elicitation and revised BBN uncertainty analysis, the number of households experiencing food insecurity and the severity of impacts are likely to increase because of the expected sizeable increases in median food prices after Brexit, evaluated here.  The effects are likely to fall disproportionately on already impoverished regions within the UK. Higher food price increases are much more likely than lower rises and, towards the upper reaches of our uncertainty distributions, these would entail very severe impacts.  Research showing a low food budget leads to increasingly poor diet suggests that demand for health services in both the short and longer term is likely to increase simply due to the effects of food insecurity on the incidence and management of diet-sensitive conditions. The recent modelling [10, 11] anticipating a 61% rise in demand for emergency food parcels in the last quarter of 2020 plus increase in demand for free school meals indicated an already incipient UK food insecurity crisis.

**Funding** The workshop was funded by the Warwick Global Research Priority for Food.  The study is part of work undertaken for EPSRC grant number EP/K007580/1.

**Competing interests** None declared